# Real-time update of multi-fractal analysis on dynamic time series using incremental discrete wavelet transforms

Nicolas Brodu

Department of Computer Science and Software Engineering
Concordia University, Montreal, Quebec, Canada, H3G 1M8
nicolas.brodu@free.fr

November 2005

**Abstract-** An algorithm is presented to update the multi-fractal spectrum of a time series in constant time when new data arrives. The discrete wavelet transform (DWT) of the time series is first updated for the new data value. This is done optimally in terms of sharing previous computations, in O(L) constant time, with L the number of levels of decomposition. The multi-fractal spectrum is then updated also in constant-time. New pre-computation techniques are presented to further accelerate this process. All possible $2^L$ data alignments are taken into account in the course of the incremental updates. The resulting spectrum estimate is more stable, compared to the current DWT method using only one dyadic frame, as precise, and more efficient. It is adapted for real-time on-line updates of the time series.

## 1 Introduction

The aim of this study is to improve on a previous multi-fractal analysis method so as to apply this technique in real-time. Precise multi-fractal analysis algorithms can be time consuming [1]. Faster techniques using wavelet transforms were developed [2], sometimes at the expense of precision [3]. Additionally, the non-stationarity of the time series considered leads to further refinements [4]. Among the latest developments, P. Manimaran *et al* propose using discrete wavelet transforms to estimate the scaling properties of a time series [5]. The present work consists in extending this technique so as to make it applicable incrementally, for dynamic time series were new data are appended in real-time, with additional stability and very good performance.

The next section introduces an incremental DWT implementation that makes use of the locality of this transformation and previous computations, so as to achieve an O(L) constant-time update as new data arrives. Section 3 describes how to make use of this algorithm to update a multi-fractal analysis also in constant-time. Details are provided for the different parts of the multi-fractal spectrum estimation. Several optional improvements are also presented. Comparison and experiments are provided in section 4, with comments and guidelines how to use this algorithm in practice. Section 5 concludes on this presentation.

## 2 Incremental DWT updates

### 2.1 Presentation of the algorithm

Discrete wavelet transform can be seen as a successive removal of details from the data, each step removing details at a given scale. The smoothed-down versions of the data are obtained by applying an averaging (low-pass) filter recursively. The details for each level of decomposition are obtained by applying the complementary (high pass) filter to the previous level averaged version, starting with the initial data. Reconstruction is achieved by the reverse process: adding the missing details to each smoothed-down level of decomposition to restore the previous level, until the original data is reconstructed. This is only a simple way to view the effect of the operations practiced on the data. It does not pretend to be complete; it's a short summary for the needs of this presentation. More information can be found in the literature on wavelet transforms (see for example [6] for pointers). The main point for this study is to emphasize:
- The recursive nature of the transform. This was presented in the previous paragraph.
- The local nature of the transform. Each filter usually having finite length, only the surrounding data of a given point is necessary to obtain the DWT.

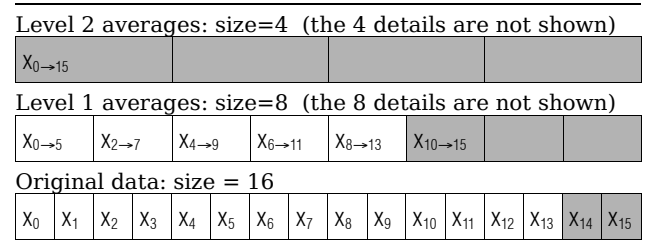

| Level 2 averages: size=4  (the 4 details are not shown) | | | |
|---|---|---|---|
| $X_{0 \to 15}$ | | | |

| Level 1 averages: size=8  (the 8 details are not shown) | | | | | | | |
|---|---|---|---|---|---|---|---|
| $X_{0 \to 5}$ | $X_{2 \to 7}$ | $X_{4 \to 9}$ | $X_{6 \to 11}$ | $X_{8 \to 13}$ | $X_{10 \to 15}$ | | |

| Original data: size = 16 | | | | | | | | | | | | | | | |
|---|---|---|---|---|---|---|---|---|---|---|---|---|---|---|---|
| $X_0$ | $X_1$ | $X_2$ | $X_3$ | $X_4$ | $X_5$ | $X_6$ | $X_7$ | $X_8$ | $X_9$ | $X_{10}$ | $X_{11}$ | $X_{12}$ | $X_{13}$ | $X_{14}$ | $X_{15}$ |

This figure shows the general form of the levels of decomposition. A low-pass filter of length 6 was applied to the data. Only the averaged data is represented. The full level of decomposition corresponds to the results of both filters: high-pass (details) and low-pass (averages). The high pass data has an identical structure.
Indices in the levels 1 and 2 indicate which of the original data were used to build this particular smoothed-down average. Grayed cells corresponds to the data that would be necessary for the reconstruction of $X_{14}$ and $X_{15}$ (in addition to the corresponding details).
**Figure 1: Smoothed-down levels of decomposition**

Figure 1 is a representation of the levels of decomposition. Each level average has a spacial resolution of twice the previous level average: the missing half of the information is contained in the details, not represented in this diagram. 2 original data are necessary to produce one average (with doubled spacial resolution) and one detail.

For a dynamic time series, where new values are appended as they become available, there is no preferred way of pairing the elements for building level 1. Figure 1 shows a pairing starting with even elements, but Figure 2 shows both alternative pairings.

| Level 1 with an even pairing | | | | | | | | |
|---|---|---|---|---|---|---|---|---|
| $X_{0 \to 5}$ | $X_{2 \to 7}$ | $X_{4 \to 9}$ | $X_{6 \to 11}$ | $X_{8 \to 13}$ | $X_{10 \to 15}$ | | | |

| Original data | | | | | | | | | | | | | | | | |
|---|---|---|---|---|---|---|---|---|---|---|---|---|---|---|---|---|
| $X_0$ | $X_1$ | $X_2$ | $X_3$ | $X_4$ | $X_5$ | $X_6$ | $X_7$ | $X_8$ | $X_9$ | $X_{10}$ | $X_{11}$ | $X_{12}$ | $X_{13}$ | $X_{14}$ | $X_{15}$ | $X_{16}$ |

| Level 1 with an odd pairing | | | | | | | | |
|---|---|---|---|---|---|---|---|---|
| | $X_{1 \to 6}$ | $X_{3 \to 8}$ | $X_{5 \to 10}$ | $X_{7 \to 12}$ | $X_{9 \to 14}$ | $X_{11 \to 16}$ | | |

A new point $X_{16}$ arrives. It cannot be paired with $X_{15}$ in the current even pairing scheme. However, using the alternative odd pairing, $X_{16}$ can be taken into account immediately.

**Figure 2: Smoothed-down levels of decomposition**

When $X_{16}$ arrives, it can be paired immediately with $X_{15}$ in the odd pairing scheme to produce $X_{11 \to 16}$. This allows to re-use the odd level 1 up to $X_{9 \to 14}$ without recomputation. Similarly, when $X_{17}$ arrives, switching back to the even pairing scheme allows to take it into account: a new point is added to the even level 1. Recursively, by keeping all $2^\lambda$ frames for each level $\lambda$, a new point can be taken into account immediately.

Since the spacial resolution decreases exponentially at each level, keeping all $2^\lambda$ frames gives back a linear memory requirement, not an exponential one. O(2LN) memory is necessary to store all the alternative frames, with L levels of decomposition, for N data points. Figure 3 sketches how this memory is used. Section 2.2 gives more information, and why much less than 2LN is actually sufficient in practice.

| Level 2 averages, size = 4 for each of the 4 frames | | | |
|---|---|---|---|
| even / even | even / odd | odd / even | odd / odd |

| Level 1 averages, size = 8 for each of the 2 frames | |
|---|---|
| even frame | odd frame |

| Original data: size = 16 | | | | | | | | | | | | | | | |
|---|---|---|---|---|---|---|---|---|---|---|---|---|---|---|---|
| $X_0$ | $X_1$ | $X_2$ | $X_3$ | $X_4$ | $X_5$ | $X_6$ | $X_7$ | $X_8$ | $X_9$ | $X_{10}$ | $X_{11}$ | $X_{12}$ | $X_{13}$ | $X_{14}$ | $X_{15}$ |

Details have the same structure as the averages. Counting both details and averages, the memory necessary to hold all the alternative frames is O(2LN). Another N would be needed to store the data itself.

**Figure 3: Frames of decomposition**

The incremental DWT update algorithm consists in the following steps:
1. Switch the next level to its alternative frame.
2. Compute the new average and detail for the next level using the new data point.
3. Recurse, using the new average as a new data for the next level.

Using cyclic buffers allows to simultaneously remove the older points without copy, as new data is added. This algorithm is very efficient, requiring only O(L) wavelet filter convolutions per update, independently of the data size N. After the update, the current set of level parity buffers holds the complete, up to date, decomposition.

An alternative way to look at this algorithm is to consider that all $2^L$ possible dyadic data frames are processed simultaneously. The structure presented in figure 3 corresponds to sharing all common computations between any 2 given frames. Conversely, a given parity buffer at level $\lambda$ is shared by $2^{L-\lambda}$ frames.

## 2.2 Practical considerations

Figure 1 shows the data necessary for the decomposition and reconstruction of particular higher level averages. As is expected in any frequency decomposition, low frequency components (corresponding to higher levels in the DWT) cannot be computed immediately. For wavelet decompositions applied to static data, like images, one can cope with this problem by considering the data periodic, or using constant padding at the ends, for example. Care must then be taken in the interpretation, as the data extension technique like roundup or padding may introduce non-negligible spurious frequencies. For a dynamic time series, such techniques are even less adequate.

When combined with the multi-fractal formalism, where discontinuities are measured, any erroneous frequency is simply not acceptable. In their presentation [5], P. Manimaran et al. compute the decompositions using constant paddings, but then discard the spurious coefficients before proceeding to the multi-fractal analysis. While this works for a static one-time analysis of data, computing and discarding coefficients is not acceptable for the real-time analysis of dynamic time series.

The solution is to accept the inevitable delay in capturing the low frequencies, and only compute the multi-fractal spectrum over the range of data that can be perfectly reconstructed. As is shown in Figure 1, for a filter of length w[1], there are (w-2) data that cannot be taken into account to compute the next level, for each level. Similarly, (w-2) data can't be reconstructed from the previous level. If the highest level L of decomposition is to hold $n_L$ data, enough to make significant statistics, then:
- $(2^L-1)(w-2)$ is the aforementioned delay to capture the low frequencies. This is also the number of oldest data that can't be reconstructed.
- $N = 2^L n_L + (2^L-1)(w-2)$ data is necessary to compute $n_L$ highest level coefficients.
- Only the most recent $R = 2^L n_L - (2^L-1)(w-2)$ out of N data is actually reconstructible.

All 3 parameters $n_L$, L, and w, are used here to compute the necessary buffer size N. This contrasts with other methods considering buffers of size $2^L$ exactly, in which the edge problem requires careful attention. For this algorithm, higher levels hold only the necessary data for reconstruction. This explains why less than 2LN memory is actually needed in section 2.1.

As aforementioned, further processing using the DWT to compute the multi-fractal spectrum is applied only to the R data that can be reconstructed.

---

1 This text assumes the reconstruction and analyzing wavelet filters have the same length for simplicity.

# 3 Constant-time update of the multi-fractal analysis

## 3.1 Presentation of the computation of the multi-fractal spectrum from the DWT

The method used in this study is based on [5]. It has been made incremental, and improved by sharing computations between updates, using a similar technique as presented in the previous part of this article. This section is a short summary of the spectrum estimation described in [5], which will serve as the basis for the algorithm that will be presented in the next section, together with additional improvements.

The first step is to extract the fluctuations of the data at a given scale from the DWT. To do this, a reconstruction $D(\lambda)$ is done without the details up to the desired level $\lambda$, and the result of that reconstruction is subtracted from the original data $D(0)$. What remains therefore consists in the original data minus its average up to level $\lambda$. This corresponds to the fluctuations $F(\lambda)$ of the data, at the scale $s=2^\lambda$ corresponding to the level $\lambda$. Note that $F(\lambda)=D(\lambda)-D(0)$ is a vector of size at most the number of reconstructible data up to level $\lambda$. For consistency, F is considered here only on the R data that can be reconstructed up to the maximum level (see section 2.2).

Once the fluctuations are extracted, it is possible to check whether their exponents exhibit a power-law scaling. The q-exponent average of $F(\lambda)$ is defined as:

$$f(\lambda,q) = \left(\frac{1}{R}\sum_{k=1}^{R}|F(\lambda)_k|^q\right)^{\frac{1}{q}} \qquad \text{Eq. 1}$$

There are L values of $f(\lambda,q)$. For each q, the h(q) exponent of the multi-fractal spectrum can be extracted from the power-law scaling of the $f(\lambda,q)$, provided such a scaling is observed. In practice, estimating h(q) amounts to an exponential data fitting between $f(\lambda,q)$ and $s^{h(q)}$ using these L values. This fitting will be described in section 3.3.

## 3.2 Incremental version

The algorithm presented in the previous section represents global statistics on all the data. This is one of the major interest of multi-fractal analysis: it summarize in a few values the behavior of the whole series. The goal is now to update these statistics efficiently as new data is presented.

The first step is to update the values of $f(\lambda,q)$, which amounts to updating the $F(\lambda)$ vector and the power sums:

$$p(\lambda,q) = \sum_{k=1}^{R}|F(\lambda)_k|^q \qquad \text{Eq. 2}$$

Unfortunately, adding a new point to the time series modifies the current decomposition frame, as described in section 2.1. Reconstruction of the data is therefore changed: The $F(\lambda)_k$ values for the series with the new point cannot be deduced from the previous $F(\lambda)_k$ values of the series before adding the new point. A naive algorithm that would add the reconstruction for the last point to the power sum in Eq. 2 is not satisfying because it would mix inconsistent terms from different frames.

The proposed solution consists in maintaining the power sums for each of the $2^L$ dyadic frames, and share the common computations between any 2 given frames, as for the DWT updates. Each level parity buffer in the structure introduced by figure 3 has now an additional field corresponding to the $p(\lambda,q)$ values for that buffer parity assignment. As before, the $p(\lambda,q)$ values for a buffer at level $\lambda$ are shared by $2^{L-\lambda}$ frames.

At this point, the next step is to update the power sums corresponding to the current decomposition. Figure 4 shows an example of power sum update for the first level.

Level 1 for the same parity pairing as the new data

| $X_{0\to5}$ | $X_{2\to7}$ | $X_{4\to9}$ | $X_{6\to11}$ | $X_{8\to13}$ | $X_{10\to15}$ |

Original data: $X_{14}$ and $X_{15}$ are new in this parity pairing

| $X_0$ | $X_1$ | $X_2$ | $X_3$ | $X_4$ | $X_5$ | $X_6$ | $X_7$ | $X_8$ | $X_9$ | $X_{10}$ | $X_{11}$ | $X_{12}$ | $X_{13}$ | $X_{14}$ | $X_{15}$ |

When new data $X_{14}$ and $X_{15}$ is added, after 2 frame switches, the first level frame is the same. Grayed background values are those already included in the power sum $p(1,q)$ for the even parity buffer: they could be reconstructed with the available data before the update (thick borders indicate the most recent data that could be reconstructed). After the new data is added, $p(1,q)$ can be updated with the fluctuations for $X_{10}$ and $X_{11}$, thanks to $X_{10\to15}$ which is now available.

**Figure 4: Data involved in a power sums update**

## 3.3 Updating the fluctuations

The direct application of the definition of the fluctuations $F(\lambda)$ implies the reconstruction without details of all the levels below the current level $\lambda$, and this has to be done for each of the L levels. While this would be a constant-time operation, there is another and faster way to compute the fluctuations for the $2^\lambda$ data points involved in the update of $F(\lambda)$.

Indeed, a wavelet filter convolution is a linear combinations of data values. So are then compositions of wavelet filters. In particular, the reconstruction $D(\lambda)$ without details up to level $\lambda$ is a linear combination of coefficients at level $\lambda$, since all details of lower levels are removed by definition. Since $D(\lambda-1)$ and $D(\lambda)$ are reconstructions without details up to levels $\lambda-1$ and $\lambda$ respectively, then the details at level $\lambda$ contain all the information for their difference. The same is also true for $F(\lambda) = D(\lambda) - D(0)$ and $F(\lambda-1) = D(\lambda-1) - D(0)$.

Therefore, it is possible to find a formula that gives $F(\lambda)$ in terms of $F(\lambda-1)$ plus a linear combination (a filter) of the details d at level $\lambda$:

$$F(\lambda) = F(\lambda-1) + \sum_{k=0}^{m-1} \alpha_k d_{k+i} \qquad \text{Eq. 3}$$

In Eq 3, the unknown parameters are m, the filter length, i, the index of the first detail d to apply the filter to, and the filter coefficients $\alpha$. All these parameters can be pre-computed by careful examination of the reconstruction process.

The grayed cells in Figure 1 show reconstruction filters of length 4 for level 2, length 3 for level 1, and that are applicable to the data at position 14 and 15. Note that $X_{10\to15}$ in level 1 is odd-indexed, and $X_{8\to13}$ could also be computed from the same details of level 2, but with different $\alpha$. Actually, there are exactly $2^\lambda$ different filters for each level $\lambda$: Data that are $2^\lambda$ apart share the same parity assignments up to level $\lambda$.

The algorithm to find α, i, and m for each of the $2^\lambda$ data index is as follow:
- Pair the data using the current parity assignment. This gives $i_{min}(0)$ and $i_{max}(0)$, for example 14 and 15 in the previous example. This will also give 2 out of the $2^\lambda$ filters.
- Recursively compute $i_{min}(\lambda+1) = \text{floor}(i_{min}(\lambda)/2) - w/2 + 1$ and $i_{max}(\lambda+1) = \text{floor}(i_{max}(\lambda)/2)$, with floor the round-down operation and w the size of the wavelet reconstruction filter as in section 2.2.
- This gives $m = i_{max} - i_{min} + 1$ for these specific data index and level.
- Set the detail $d_{k+i\_min}$ to 1 and all other details to 0. Thanks to the linear nature of wavelet convolutions, reconstructing down to level 0 (the original data level) gives the 2 coefficients $\alpha_k$ for the 2 filters corresponding to the initial index $i_{min}(0)$ and $i_{max}(0)$.
- Repeat the previous step for all k=0..m-1 to get the full filters for level λ at the chosen data positions.
- Loop from the first step for all $2^\lambda$ data positions.
- Repeat the whole procedure for all levels λ=1..L.

This pre-computation is only dependent on a given wavelet and number of levels. It can be shared amongst analyzers processing different time series, for example. Once available, the filters can be used directly in Eq. 3. The power sums can now be updated in a much more efficient way.

Starting from level 0 (the original data), an array of size $2^L$ values is initialized to 0. This array will contain the F(λ) values, as shown in Figure 5. It is passed from level to level, and updated according to Eq. 3. At this point, the previous $2^L-2^\lambda$ values of F(λ) could be stored in memory and re-used, as for the power sums, or recomputed. This choice largely depends on the application: the memory requirement to store past values is $2^L-2^\lambda$ for each of the $2^\lambda$ parity buffers of a given level, which adds up very quickly. Whereas recomputing the $2^L-2^\lambda$ values directly from the data thanks to the filters is fast, though not as fast as a copying past computations. Either way is faster than the reconstruction down to data level suggested by the direct application of the F(λ) definition.

F(4), assuming $X_{15}$ is the latest reconstructible data.

| $F_0$ | $F_1$ | $F_2$ | $F_3$ | $F_4$ | $F_5$ | $F_6$ | $F_7$ | $F_8$ | $F_9$ | $F_{10}$ | $F_{11}$ | $F_{12}$ | $F_{13}$ | $F_{14}$ | $F_{15}$ |
|---|---|---|---|---|---|---|---|---|---|---|---|---|---|---|---|

F(3), assuming $X_{15}$ is the latest reconstructible data.

| $F_0$ | $F_1$ | $F_2$ | $F_3$ | $F_4$ | $F_5$ | $F_6$ | $F_7$ | $F_8$ | $F_9$ | $F_{10}$ | $F_{11}$ | $F_{12}$ | $F_{13}$ | $F_{14}$ | $F_{15}$ |
|---|---|---|---|---|---|---|---|---|---|---|---|---|---|---|---|

F(2), assuming $X_{15}$ is the latest reconstructible data.

| $F_0$ | $F_1$ | $F_2$ | $F_3$ | $F_4$ | $F_5$ | $F_6$ | $F_7$ | $F_8$ | $F_9$ | $F_{10}$ | $F_{11}$ | $F_{12}$ | $F_{13}$ | $F_{14}$ | $F_{15}$ |
|---|---|---|---|---|---|---|---|---|---|---|---|---|---|---|---|

F(1), assuming $X_{15}$ is the latest reconstructible data.

| $F_0$ | $F_1$ | $F_2$ | $F_3$ | $F_4$ | $F_5$ | $F_6$ | $F_7$ | $F_8$ | $F_9$ | $F_{10}$ | $F_{11}$ | $F_{12}$ | $F_{13}$ | $F_{14}$ | $F_{15}$ |
|---|---|---|---|---|---|---|---|---|---|---|---|---|---|---|---|

After new data has added, let's suppose that $X_{15}$ is the latest reconstructible data. In order to update the power sums in each level λ, the grayed background F(λ) values have to be computed. The white background values could either be recomputed or reused from past updates, depending on the application requirements.

**Figure 5: Updating the F(λ) values**

Each level then computes only the $2^\lambda$ last values of $|F(\lambda)|^q$ in order to update the power sums (the grayed cells in Figure 5). The first $2^L-2^\lambda$ values of F(λ) are not exponentiated, they are just needed to build further levels up.

### 3.4 Taking care of old data

Up to this point only new data was considered, but older values should also be removed. Unfortunately, it is not possible to recompute the past $|F(\lambda)|^q$ values, since the corresponding data was precisely discarded in the cyclic buffers. For stationary series a solution would be to make statistics from the first point of the time series and ignore the problem. Without this assumption, the older points should be removed.

The solution is to compute the $2^\lambda$ first F(λ) values just before updating the DWT with the new data, and remove these from the corresponding power sums. After a whole cycle of the $2^\lambda$ frames, the same parity alignment is reached again for the parity buffers at level λ. The power sum is ready to be updated with the new data, without the old points. The algorithm now looks like:
- Compute the $2^\lambda$ first values of $|F(\lambda)|^q$ for the current frame and associated set of parity buffers, for each level λ. This is done with the current data and the adequate filters according to Eq 3.
- Prepare the power sums of the current frame for a future update by removing these old values.
- Update the DWT incrementally as described in section 2. This changes the current frame and parity buffers.
- Compute the $2^\lambda$ last values of $|F(\lambda)|^q$ for the new current frame. Update the power sums with the new data.

At the end of these steps, the power sums of Eq. 2 always have an up to date value. The multi-fractal spectrum can now be derived.

### 3.5 Spectrum computation

As mentioned in section 3.1, the h(q) exponents for multi-fractal analysis are estimated from the power-law scaling of the f(λ,q) values. For each scale $s=2^\lambda$, the goal is to find the best h(q) that fits:

$$f(\lambda,q) \propto s^{h(q)} \qquad \text{Eq. 4}$$

with ∝ meaning "proportional to". Eq 4. can be rewritten:

$$\log_2 f(\lambda,q) \approx h(q) \cdot \lambda + C_1$$
$$\log_2 p(\lambda,q) \approx h(q) \cdot q\lambda + C_2 \qquad \text{Eq. 5}$$

with $C_1$ a constant independent of λ, and $C_2=qC_1+\log_2 R$ according to Eq 2 and 3, so $C_2$ is also independent of λ.

The simplest way of proceeding consists in a least square estimation for h(q) using the L power sum logarithms. But for exponential fitting this would introduce a bias: lower scales would be given a higher importance than higher scales in the estimation. Weighted least square fitting allows to restore the relative influence of each scale by choosing adequate weights. A better way to estimate the h(q) values is thus to minimize:

$$\sum_{\lambda=1}^{L} w_\lambda \left( \log_2 p(\lambda, q) - (h(q) \cdot q\lambda + C_2) \right)^2 \quad \text{Eq. 6}$$

The reader is invited to consult [7] for example, for the derivation of the weighted least square estimate solution. When rewriting that solution with the parameters of Eq. 6, the h(q) are given in analytical form by:

$$h(q) = \frac{1}{q} \sum_{\lambda=1}^{L} \mu_\lambda \log_2 p(\lambda, q) \quad \text{Eq. 7}$$

with:

$$\mu_k = \frac{w_k \left( k \sum_{\lambda=1}^{L} w_\lambda - \sum_{\lambda=1}^{L} w_\lambda \lambda \right)}{\sum_{\lambda=1}^{L} w_\lambda \cdot \sum_{\lambda=1}^{L} w_\lambda \lambda^2 - \left( \sum_{\lambda=1}^{L} w_\lambda \lambda \right)^2} \quad \text{Eq. 8}$$

The unweighted least square fitting corresponds to all weights $w_\lambda$ being equal.

The solution given by Eq. 7 and Eq. 8 has been rewritten in such a way as to clearly exhibit a $\mu_\lambda$ term: this way, $\mu_\lambda$ deliberately appears as a weight in Eq. 7. Using Eq. 8 it is possible to pre-compute $\mu_\lambda / q$ for each q, $\lambda$. All there is to do in real-time is then to add the weighted logarithms as in Eq. 7, which can be done very efficiently[2].

The question is now to find proper weights so all scales are given approximately equal influence. This is achieved by using an exponential relation for the weights, ideally proportional to the fittest exponential itself. For the fitting of $y = A \cdot 2^{Bx}$, one choice [7] could to use the y values as weights: they are supposedly close to the fittest exponential[3]. Unfortunately, for a fast computation of the h(q), it is imperative that the $\mu_\lambda$ do not depend on the power sums, otherwise their pre-computation would not be possible.

What is proposed in this algorithm is to choose an exponential scale that corresponds to a classical brownian motion: integrated white noise [8], a mono-fractal with Hurst exponent 0.5. Without any a priori knowledge on the data (hence without y values), no guess can be made about long term dependence of the data on previous values. With H=0.5, no assumption is made on persistence (H>0.5) or anti-persistence (H<0.5) [9]. On the one hand, this isn't nearly as good as the y values weighting scheme aforementioned. On the other hand, this is an exponential scale that gives better results than an unweighted scheme at no cost, since it allows pre-computations. The proposed weights are thus given by:

$$w_\lambda = s^{0.5} = \sqrt{2^\lambda} \quad \text{Eq. 9}$$

---

[2] By chance, personal computers (amongst others) with the x86 instruction set (x87 FPU) can process "y·log₂x" in a single assembly instruction. This can be used to compute Eq 7 very efficiently.

[3] There are still problems with this approach: a spurious y value used as weight would precisely enhance the relative influence of this spurious point. An iterative solution may be to use a first exponential estimate to build a new set of weights, and so on, until convergence. But this would be too expensive computationally for this algorithm, and not justified in regards of the weight experiment results, section 4.2.

Depending on the user application, other choices may be more adequate. Many natural phenomena exhibit "pink noise" [10]: power-law scaling proportional to the frequency inverse. Integrated pink noise has a Hurst exponent of 1.0. A choice of $w_\lambda = 2^\lambda$ may be more appropriate for such occasions. The proposed publicly available implementation of this algorithm (see Appendix 1) default to Eq 9, but accepts user-defined weights as optional parameters.

The power-sum updating routine presented in 3.2 now includes an additional step. As the p(λ,q) values are updated, so are the h(q): the weighed logarithm of p(λ,q) is added for each level. When the highest level is reached, the new h(q) are ready.

This completes the incremental multi-fractal algorithm presentation. The DWT was updated for the new data, re-using previous computations from the same dyadic frame. The power sums were updated quickly, thanks to a filter that allows direct computation of the F(λ) from the data. Finally, a fast method was given to obtain an estimate of the h(q) from the power sums, by pre-computing the factors for a weighted exponential fit.

### 3.6 Optional context-dependent improvements

So far, each of the $2^L$ frames have their own power series and h(q) estimates. Data is shared optimally amongst the different frames, but the final power sums p(λ,q) are still nonetheless maintained for each level parity buffer.

An optional improvement may thus be to average the h(q) values over all $2^L$ latest frames. For direct average computations, all there is to do is dividing each pre-computed $\mu_\lambda$ by $2^L$ (see Eq 7). This way, the new $h(q)/2^L$ is obtained without modification to the previous algorithm. Updating the average is immediate. A $2^L$ history of $h(q)/2^L$ is maintained for each q. As a new $h(q)/2^L$ is computed, it is added to the average and the history. The oldest $h(q)/2^L$ is removed from the history and subtracted from the average.

The advantage of averaging over the $2^L$ frames is to reduce the frame to frame variations in the estimates (see experiment section 4.2). For stationary time series especially, this brings more stable and reliable results.

The drawback is the dilution of the influence of the newest data point. For non-stationary time series, averaging over $2^L$ frames enlarges the number of data the latest estimate is based on by $2^L-1$. However, depending on the context, the benefits of having estimates less sensitive to frame to frame variations may out-weight this drawback. The public implementation provided with this article (see Appendix 1) considers averaging as an optional operation that is active by default, but can be deactivated depending on the user application.

Another optional improvement for stationary series concerns the possibility of keeping all points in the power sums, instead of removing the old ones (with the additional benefit of reduced computation time). This is an improvement only for stationary time series, where more data means better estimates. However, this is not desirable for non-stationary series, where older points have different properties than the newer ones. Hence this possibility is also kept as an option, unset by default, in the proposed implementation aforementioned.

## 4 Experiments

### 4.1 Numerical stability

One potential source of numerical roundup errors lies in the incremental update of the power sums (Eq 2). When adding a large number of values, while subtracting older ones at the same time, floating point operations may cause a drift in the numerical result vs the analytical one. The final values obtained with this incremental algorithm may therefore be different from the non-incremental version.

To investigate this effect, a non-incremental version of the algorithm is also provided. This non-incremental version was also developed for the initialization phase of the incremental algorithm, so as to provide the first power sums. The alternative would be to give h(q) estimates during the initialization phase with only the first levels of decomposition, so long as there is not enough data. Since this would require the computation of adequate $\mu_\lambda$ weights for the initialization phase, a non-incremental call when enough data is available was preferred.

Thanks to the non-incremental version, it is possible to measure the floating point drift induced by Eq 2. An experiment was set up to monitor the average number of iterations necessary to notice a $10^{-6}$ error in the h(q) estimates between the incremental and the non-incremental version. In practice, the time it takes to notice a difference depends on the time series values. Statistics are made on 30 independent runs to quantify the deviation and derive the error bars if needed.

This experiment is set up using Brownian motion with H=0.5. It is reproducible as a test program coming with the source code. An analyzer with 5 levels, q exponents from -10 to +10, and 30 points at the highest scale was used on the random source. No averaging is done on consecutive frames, and past data are discarded: Averaging over the $2^L$ frames would reduce the influence of an imprecision on one of the parity-aligned independent power sums. Removing past data causes twice as many changed terms per update, compared to just adding the new values.

The results are that even after 10 million updates, no drift was found at $10^{-6}$ precision for all positive q exponents, and for q=-1, in any of the independent runs. q=-2 exhibits a $10^{-6}$ difference after the order of a hundred thousand updates (with a large standard deviation of 70 thousand steps). Nevertheless, no drift above $1.5 \cdot 10^{-4}$ was observed after the 10 million updates in each of the runs for q=-2. Larger negative q exponents are more sensitive. The largest negative q diverge at $10^{-6}$ after a few thousand iterations only, and may exhibit drifts the order of $10^{-1}$ or $10^{-2}$ after the 10 million updates. However, the largest negative q are also the most unreliable ones in the base version of the wavelet transform.

Given that the inherent precision of a wavelet transform analysis is seldom below $10^{-3}$ for positive q anyway (see next section), less precise for negative q, and given that it would take at least more than 10 million updates to see a numerical drift above that inherent precision: it is safe to assume that the incremental algorithm is sufficiently numerically stable for most uses, compared to the non-incremental one.

### 4.2 Precision

The goal of this section is to measure the effect of the new options brought by this algorithm: influence of keeping or discarding old points and of the number $n_L$ of data at the highest scale, influence of the weighting scheme, and influence of averaging over the $2^L$ frames. The precision inherent to the discrete wavelet transform method, as well as the influence of parameters like the choice of the wavelet, were already discussed in [5].

Experiments were done using the Daubechies Wavelet with 6 coefficients. A mono-fractal Brownian motion random source with theoretical H=0.5, and an integrated pink noise with H=1.0, are both investigated. The white noise is generated using the Mersenne Twister algorithm [11] and the pink noise is generated by the Voss - McCartney algorithm (see Appendix 1). All these implementations are provided with the source code so the experiments are easily reproducible.

Each experiment was done with the correct weighting scheme respectively corresponding to the theoretical H=0.5 and H=1.0: this avoids the exponential data fit bias toward small scales explained in Section 3.3. The experiments were then repeated with basic least square fitting, with all weights equal to 1.0, for comparison. Overall statistics are made on 30 independent runs. The other effect that is investigated is the averaging of the h(q) over the $2^L$ frames. Frame-to-frame variations are measured within each run, in order to get standard deviation from the frame average.

Results are given in Appendix 2. Figure 7 shows the overall precision experiment for integrated pink noise, when keeping old values: statistics are made on the whole series, not just the latest data. 3 different configurations are selected, each corresponding to increased number of levels of decomposition. The results are insensitive to the number $n_L$ of data at the highest level since old values are kept (this was verified experimentally with $n_L$=10, 30 and 50 giving exactly the same results).

As expected, the overall precision increases with each configuration. Even the low-quality L=4 configuration produces reasonable results, with a better than $5 \cdot 10^{-2}$ precision on average, for a theoretical H=1.0. However, the run-to-run standard deviation is too high for reliable estimates. On a given specific run, which is the case for on-line time series updating, the observed precision may fluctuate too much from the average, especially for large negative q values. With the medium L=7 configuration the precision is slightly increased, with less than $4 \cdot 10^{-2}$ error on the whole spectrum. But more importantly, the run-to-run variation is now acceptable. The higher-quality L=10 configuration further improves the result, though not as much as the L=7 one did over the L=4 configuration. Consequently, there is little interest in increasing the number of levels above a threshold value, especially when considering the performance cost this entails (see next section).

Figure 8 shows the effect of the weighting scheme, with the same setup. Once again the effect is more important for the low-precision configuration. For the medium and high ones, the gain is noticeable, though it remains just above the precision level aforementioned. Results using the Brownian motion with H=0.5 instead of the integrated pink noise with H=1.0 are even less sensitive to the weighting scheme. This implies the choice of the weights is not crucial, and the default configuration of the weights should be adapted to most uses. However, since this algorithm implements weighting at no cost, thanks to the pre-computation of the exponential fit coefficients, the correct weighting scheme might as well be used whenever possible.

The third part of the precision experiment concerns the frame to frame deviations. Results are shown for the Brownian motion. The old points are discarded, and as a result the number of data $n_L$ becomes significant (see figure 9). The frame averaged results are given first. The observed precision for L=7 and 10 is the same order as for the previously presented overall series statistics (with old points kept): precision the order of $3.10^{-2}$. For L=4, $n_L$=10, the results are worse, but this is due to the low $n_L$ value: when increasing it to L=4, $n_L$=50, another experiment shows the average precision goes back to $6.10^{-2}$, which is comparable to the $5.10^{-2}$ in the whole series experiment.

The run-to-run deviations are not shown: they are similar to the previous experiment (with the same note on the unreliability of L=4). Consequently, provided the number of data at the higher level $n_L$ is not too low, results are comparable to the whole series version. This seems logical, as the series keeps the same properties through time: discarding the old points has little effect so long as there are enough data to average on. For a series with properties changing through time, removing old points would be necessary.

The frame-to-frame deviations are shown next to the frame averages. The low-quality L=4, $n_L$=10 exhibits the worst variations, with unusable results. L=7, $n_L$=30, gives stable frame to frame results for positive q, but unstable ones for negative q. L=10, $n_L$=50 exhibits frame deviations below the precision level. Interestingly, in another experiment, increasing $n_L$ for L=4 and L=7 doesn't significantly improve the frame to frame variations, unlike the run-to-run variations. Therefore, whereas frame averaging can be omitted for L=10 given these results, it is really beneficial to medium-quality configurations by increasing the stability in the estimates.

### 4.3 Performance

The performance experiment concerns the relative speed of the incremental algorithm compared to the non-incremental one. The absolute timings and the relative gain of the incremental version are given for different number of data and number of levels. All computations are done with the Daubechies wavelet with 6 coefficients. These experiments measure the time it takes to update a multi-fractal analysis by adding a new data value and removing an old one. Timings when keeping the old points are slightly above half these results. Absolute timings are dependent on a specific hardware configuration, so 2 configurations are tested. Some optimizations like the pre-computation of the filter (see section 3.3) and of the exponential data fit (see section 3.5) benefit to both the incremental and the non-incremental versions of the algorithm.

The computation time is independent from the data values, but other effects introduce some variance: the OS multitasking, different branching predictions for each update on the CPU, the resolution of the timer used for measurement, etc. In practice, it's necessary to average the experiments over large a number of updates. Figure 6 shows the result.

| L | | $n_L$=10 | | $n_L$=30 | | $n_L$=50 | |
|---|---|---|---|---|---|---|---|
| | | conf. 1 | conf. 2 | conf. 1 | conf. 2 | conf. 1 | conf. 2 |
| 4 | N | 47.3 (1.08) | 45.8 * (27.6) | 153 (2.36) | 119 * (24.7) | 257 (2.22) | 185 * (11.2) |
| | I | 17.4 (0.859) | 9.09 * (13.9) | 17.5 (0.565) | 13.1 * (15.0) | 17.7 (0.677) | 12.1 * (14.8) |
| | R | 2.73 | 5.04 * | 8.74 | 9.04 * | 14.5 | 15.3 * |
| 7 | N | 512 (1.89) | 366 (4.69) | 2000 (4.66) | 1425 (8.93) | 3520 (5.66) | 2489 (14.2) |
| | I | 104 (1.04) | 73.2 (2.33) | 104 (1.04) | 73.7 (1.40) | 104 (1.12) | 74.3 (2.10) |
| | R | 4.94 | 5.00 | 19.3 | 19.3 | 34.0 | 33.5 |
| 10 | N | 9480 (362) | 5190 (24.5) | 36700 (1630) | 20800 (111) | 61500 (2900) | 36600 (91.8) |
| | I | 2520 (114) | 1080 (12.7) | 2550 (128) | 1080 (20.3) | 2540 (127) | 1080 (15.0) |
| | R | 3.76 | 4.79 | 14.4 | 19.2 | 24.2 | 34.0 |

Legend:
| | |
|---|---|
| time (dev) | Absolute time in microseconds for one update, averaged over the whole batch of runs, with the observed standard deviation. |
| L | Number of levels |
| $n_L$ | Size of level L. See section 2.2. |
| N | Non-incremental version |
| I | Incremental version |
| R | Ratio of the mean timings N / I |
| Conf. 1 & 2 | Machine configurations (see text) |

These are measurements of the time it takes to process a new data value and to remove an old one for the multi-fractal analysis using the Daubechies wavelet with 6 coefficients. All timings are given in microseconds, rounded to 3 digits. The experiment was repeated on a Athlon 1600+/1.4 Ghz system (conf. 1) and an Athlon 3000+/1.8 Ghz (conf. 2), with optimizations turned on in each case. Results marked * are less reliable due to some measures being below the system timer precision.

**Figure 6: Performance measurements**

As expected, increasing the number of data has no significant effect on the incremental version timings (see the "I" rows in Figure 6). Consequently, the gain of the incremental algorithm over the non-incremental one increases with the data size. Similarly, the ratios depend much more on the algorithms themselves than on the machine speed. This is observed for L=7 especially. An unexpected result is the the influence of a faster machine on the ratios for large number of data. An explanation would probably be machine dependent, and the influence of all system parameters would have to be taken into account (larger memory cache, data transfer time, etc).

The effect of reducing the number of q-exponent has also been investigated. A detailed profiler analysis shows that the incremental algorithm spends only part of its time processing the $|F(\lambda)|^q$ exponentiations: there are only ($n_q$-1) multiplications to process $n_q$ positive q values, and the same number plus an additional division to process $n_q$ negative values. Reducing the number of q causes a speed gain, but not proportional one. For example, for L=7, on machine configuration 1, the incremental update time goes down from 104 to 82 microseconds when considering q=-5..5 instead of q=-10..10. As a result, this algorithm is well suited to estimate large ranges of q values.

All in all, when considering the precision results of the previous section, there seems to be little gain in increasing the number of levels above some intermediate value.

## 5 Conclusion

This new algorithm for updating a multi-fractal analysis in constant time has the same capacities, inherent precision, and properties, as the method based on discrete wavelet transform presented in [5]. However, it is much more efficient, even in its non-incremental form, thanks to the pre-computation of different crucial parts. The incremental updates add another order of magnitude of performance. Optionally averaging over all $2^L$ dyadic frames provides more stability in the estimates for medium levels of decomposition.

The incremental DWT part of this algorithm may in itself have other applications beside the estimation of multi-fractal spectra. Both the DWT and the multi-fractal parts are fully generic, they make no assumption on the data.

This algorithm is numerical stable, precise, and efficient. It is therefore applicable to a wide variety of applications requiring real time computations.

## Acknowledgments

Financial support was provided by the EADS Corporate Research Center, in cooperation with the French Ministry of Foreign Affairs.

## Appendix 1: Software availability

This project source code is available under the GNU General Public License, v2 or above. Links can be found on the author web page http://nicolas.brodu.free.fr.

All random seeds and parameters used in the experiments are included in test programs provided with the source code. The white noise random number generator is the Mersenne Twister by Makoto Matsumoto [11], http://www.math.sci.hiroshima-u.ac.jp/~m-mat/MT/emt.html.

The pink noise is generated by the Voss-McCartney algorithm http://www.firstpr.com.au/dsp/pink-noise/, but using the Mersenne twister as white noise source. These random number generators are provided with the source code.

All other algorithms and routines are my personal creation.

# Appendix 2: Precision experiment results

| | L=4 | | L=7 | | L=10 | |
|---|---|---|---|---|---|---|
| q | h(q) | dev. | h(q) | dev. | h(q) | dev. |
| 1 | 1.0108 | 0.0039 | 1.0106 | 0.0076 | 1.0060 | 0.0236 |
| 2 | 1.0091 | 0.0039 | 1.0103 | 0.0074 | 1.0031 | 0.0235 |
| 3 | 1.0083 | 0.0042 | 1.0103 | 0.0078 | 1.0008 | 0.0241 |
| 4 | 1.0086 | 0.0047 | 1.0105 | 0.0086 | 0.9983 | 0.0251 |
| 5 | 1.0096 | 0.0052 | 1.0107 | 0.0097 | 0.9954 | 0.0263 |
| 6 | 1.0112 | 0.0058 | 1.0109 | 0.0109 | 0.9921 | 0.0274 |
| 7 | 1.0132 | 0.0065 | 1.0110 | 0.0122 | 0.9886 | 0.0284 |
| 8 | 1.0155 | 0.0072 | 1.0108 | 0.0134 | 0.9850 | 0.0292 |
| 9 | 1.0179 | 0.0079 | 1.0104 | 0.0146 | 0.9814 | 0.0300 |
| 10 | 1.0202 | 0.0087 | 1.0098 | 0.0157 | 0.9780 | 0.0305 |
| -1 | 1.0205 | 0.1098 | 1.0171 | 0.0251 | 0.9995 | 0.0342 |
| -2 | 1.0417 | 0.2823 | 1.0289 | 0.0517 | 1.0015 | 0.0450 |
| -3 | 1.0436 | 0.3105 | 1.0310 | 0.0548 | 1.0020 | 0.0467 |
| -4 | 1.0439 | 0.3190 | 1.0315 | 0.0556 | 1.0022 | 0.0474 |
| -5 | 1.0438 | 0.3225 | 1.0317 | 0.0560 | 1.0024 | 0.0477 |
| -6 | 1.0436 | 0.3242 | 1.0318 | 0.0562 | 1.0024 | 0.0480 |
| -7 | 1.0435 | 0.3252 | 1.0318 | 0.0563 | 1.0025 | 0.0481 |
| -8 | 1.0434 | 0.3259 | 1.0318 | 0.0564 | 1.0025 | 0.0482 |
| -9 | 1.0433 | 0.3263 | 1.0317 | 0.0565 | 1.0025 | 0.0483 |
| -10 | 1.0432 | 0.3266 | 1.0317 | 0.0565 | 1.0025 | 0.0483 |

Results are presented here for integrated pink noise, a theoretical mono-fractal with H=1.0. The h(q) values are shown for different q, averaged over 30 independent runs. The run-to-run standard deviation is also shown. L stands for the number of levels of decomposition. The number of data $n_L$ at the highest level has no influence on these results, since all values in the time series are kept (the old points are not discarded).

**Figure 7: Precision measurements**

| | L=4 | | L=7 | | L=10 | |
|---|---|---|---|---|---|---|
| q | |Δ| no weight | dev. | |Δ| no weight | dev. | |Δ| no weight | dev. |
| 1 | 0.0025 | 0.0014 | 0.0046 | 0.0036 | 0.0163 | 0.0121 |
| 2 | 0.0013 | 0.0010 | 0.0047 | 0.0033 | 0.0168 | 0.0116 |
| 3 | 0.0026 | 0.0015 | 0.0049 | 0.0036 | 0.0175 | 0.0120 |
| 4 | 0.0040 | 0.0019 | 0.0053 | 0.0040 | 0.0183 | 0.0129 |
| 5 | 0.0049 | 0.0020 | 0.0057 | 0.0045 | 0.0194 | 0.0143 |
| 6 | 0.0053 | 0.0021 | 0.0061 | 0.0050 | 0.0212 | 0.0148 |
| 7 | 0.0054 | 0.0022 | 0.0068 | 0.0053 | 0.0230 | 0.0158 |
| 8 | 0.0052 | 0.0024 | 0.0077 | 0.0055 | 0.0246 | 0.0170 |
| 9 | 0.0050 | 0.0024 | 0.0086 | 0.0057 | 0.0263 | 0.0180 |
| 10 | 0.0048 | 0.0024 | 0.0094 | 0.0062 | 0.0279 | 0.0190 |
| -1 | 0.0747 | 0.0582 | 0.0308 | 0.0263 | 0.0295 | 0.0289 |
| -2 | 0.1841 | 0.1436 | 0.0764 | 0.0612 | 0.0580 | 0.0438 |
| -3 | 0.2018 | 0.1553 | 0.0840 | 0.0660 | 0.0631 | 0.0463 |
| -4 | 0.2072 | 0.1593 | 0.0862 | 0.0678 | 0.0647 | 0.0474 |
| -5 | 0.2096 | 0.1611 | 0.0872 | 0.0688 | 0.0653 | 0.0480 |
| -6 | 0.2108 | 0.1621 | 0.0877 | 0.0693 | 0.0657 | 0.0483 |
| -7 | 0.2116 | 0.1626 | 0.0880 | 0.0697 | 0.0659 | 0.0486 |
| -8 | 0.2121 | 0.1630 | 0.0882 | 0.0697 | 0.0660 | 0.0487 |
| -9 | 0.2125 | 0.1632 | 0.0883 | 0.0701 | 0.0660 | 0.0488 |
| -10 | 0.2128 | 0.1633 | 0.0884 | 0.0703 | 0.0661 | 0.0489 |

These results are produced by the same experiment as described in Figure 7. |Δ| means the absolute value of the observed difference between the result with the correct exponential fit weighting scheme (Figure 7), and the result with an unweighted least square fitting.

**Figure 8: Influence of the weighting scheme**

| | L=4, $n_L$=10 | | L=7, $n_L$=30 | | L=10, $n_L$=50 | |
|---|---|---|---|---|---|---|
| q | h(q) | frame dev. | h(q) | frame dev. | h(q) | frame dev. |
| 1 | 0.5649 | 0.0036 | 0.5143 | 0.0004 | 0.5043 | 0.0015 |
| 2 | 0.5538 | 0.0034 | 0.5121 | 0.0004 | 0.5036 | 0.0015 |
| 3 | 0.5423 | 0.0035 | 0.5098 | 0.0004 | 0.5024 | 0.0015 |
| 4 | 0.5320 | 0.0038 | 0.5074 | 0.0004 | 0.5010 | 0.0015 |
| 5 | 0.5235 | 0.0041 | 0.5047 | 0.0004 | 0.4994 | 0.0015 |
| 6 | 0.5168 | 0.0045 | 0.5018 | 0.0005 | 0.4975 | 0.0015 |
| 7 | 0.5117 | 0.0048 | 0.4988 | 0.0005 | 0.4955 | 0.0015 |
| 8 | 0.5077 | 0.0051 | 0.4958 | 0.0005 | 0.4933 | 0.0015 |
| 9 | 0.5046 | 0.0053 | 0.4929 | 0.0006 | 0.4911 | 0.0014 |
| 10 | 0.5023 | 0.0055 | 0.4901 | 0.0006 | 0.4888 | 0.0014 |
| -1 | 0.4947 | 0.1878 | 0.5214 | 0.0289 | 0.5110 | 0.0022 |
| -2 | 0.4338 | 0.4630 | 0.5217 | 0.1114 | 0.5173 | 0.0048 |
| -3 | 0.4185 | 0.5455 | 0.5233 | 0.1318 | 0.5176 | 0.0054 |
| -4 | 0.4081 | 0.5745 | 0.5241 | 0.1391 | 0.5178 | 0.0056 |
| -5 | 0.4127 | 0.5815 | 0.5248 | 0.1428 | 0.5179 | 0.0057 |
| -6 | 0.4080 | 0.5709 | 0.5250 | 0.1460 | 0.5180 | 0.0057 |
| -7 | 0.3915 | 0.5710 | 0.5278 | 0.1492 | 0.5182 | 0.0058 |
| -8 | 0.4290 | 0.5746 | 0.5295 | 0.1532 | 0.5183 | 0.0058 |
| -9 | 0.4669 | 0.5918 | 0.5140 | 0.1579 | 0.5184 | 0.0058 |
| -10 | 0.4566 | 0.6037 | 0.5179 | 0.1627 | 0.5187 | 0.0058 |

Results are presented for Brownian motion with incremental updates discarding the old points. The mean over all runs frame to frame deviation is shown. The run-to-run deviation of the h(q) is the same order as in figure 7.

**Figure 9: Frame-to-frame deviations**